\documentclass[aps,twocolumn,prd,superscriptaddress,nofootinbib,showpacs,letterpaper,eqsecnum]{revtex4}

\usepackage{amsmath,amssymb}
\usepackage[dvips]{graphicx}
\usepackage[export]{adjustbox}
\usepackage{longtable}
\usepackage[usenames,dvipsnames]{xcolor}
\usepackage[normalem]{ulem}
\usepackage{bm}
\usepackage{comment}
\usepackage{xcolor}
\usepackage{soul}
\usepackage{hyperref}
\usepackage[percent]{overpic}
\usepackage{overpic}

\begin{document}

\newcommand{\bea}{\begin{eqnarray}}
\newcommand{\eea}{\end{eqnarray}}
\newcommand{\E}{\mathrm{E}}
\newcommand{\Var}{\mathrm{Var}}
\newcommand{\bra}[1]{\langle #1|}
\newcommand{\ket}[1]{|#1\rangle}
\newcommand{\braket}[2]{\langle #1|#2 \rangle}
\newcommand{\mean}[2]{\langle #1 #2 \rangle}
\newcommand{\be}{\begin{equation}}
\newcommand{\ee}{\end{equation}}	
\newcommand{\ba}{\begin{eqnarray}}
\newcommand{\ea}{\end{eqnarray}}
\newcommand{\SD}[1]{{\color{magenta}#1}}
\newcommand{\rem}[1]{{\sout{#1}}}
\newcommand{\alert}[1]{\textbf{\color{red} \uwave{#1}}}
\newcommand{\Y}[1]{\textcolor{blue}{#1}}
\newcommand{\R}[1]{\textcolor{red}{#1}}
\newcommand{\B}[1]{\textcolor{black}{#1}}
\newcommand{\C}[1]{\textcolor{cyan}{#1}}
\newcommand{\db}{\color{darkblue}}
\newcommand{\huan}[1]{\textcolor{cyan}{#1}}
\newcommand{\fan}[1]{\textcolor{blue}{#1}}
\newcommand{\ac}[1]{\textcolor{cyan}{\sout{#1}}}
\newcommand{\intinfty}{\int_{-\infty}^{\infty}\!}
\newcommand{\Tr}{\mathop{\rm Tr}\nolimits}
\newcommand{\const}{\mathop{\rm const}\nolimits}
\newcommand{\Caltech}{\affiliation{Theoretical Astrophysics 350-17, California Institute of Technology, Pasadena, CA 91125, USA}}

\title{Towards an understanding of the force-free magnetosphere \\of rapidly spinning black holes}
\author{Fan Zhang}
\affiliation{Center for Cosmology and Gravitational Wave, Department of Astronomy, Beijing Normal University, Beijing 100875, China}
\affiliation{\mbox{Department of Physics, West Virginia University, PO Box 6315, Morgantown, WV 26506, USA}}
\author{Huan Yang}
\affiliation{Perimeter Institute for Theoretical Physics, Waterloo, Ontario N2L 2Y5, Canada}
\affiliation{Institute for Quantum Computing, University of Waterloo, Waterloo, Ontario N2L3G1, Canada}
\email{hyang@perimeterinstitute.ca}
\author{Luis Lehner}
\affiliation{Perimeter Institute for Theoretical Physics, Waterloo, Ontario N2L 2Y5, Canada}
\affiliation{CIFAR, Cosmology \& Gravity Program, Toronto, ON M5G 1Z8, Canada}
\email{llehner@perimeterinstitute.ca}

\begin{abstract}
The ability of a plasma surrounding spinning black holes to extract rotational energy and power energetic 
emissions has been recognized as a key astrophysical phenomenon. Important insights into the nature
of this process are obtained through the analysis of the interplay between a 
force-free magnetosphere and the black hole. This task involves solving a complicated system of equations, often
requiring complex numerical simulations.
Recent analytical attempts at tackling this problem have exploited the fact that the
near horizon region of extreme Kerr (NHEK) is endowed with an enhanced symmetry group.
We continue in this direction and show that for some conformally self-similar solutions, the NHEK force-free equations reduce to a single non-linear ordinary differential equation which is difficult to solve with
straightforward integration. We here introduce a new approach specifically tailored to this problem and
describe how one can obtain physically meaningful solutions.
\end{abstract}


\maketitle

\section{Introduction}
Energetic, highly collimated, emissions emanating from a localized central engine are observed throughout our universe. A leading model to explain the engine powering these jets involves, at a basic level, a spinning black hole feeding its rotational energy into the kinetic and thermal energy of the surrounding plasma, through a process such as the Blandford-Znajek mechanism \cite{1977MNRAS.179..433B}. While this basic picture is widely accepted, a detailed understanding of these systems remain elusive. This status of affairs is due to
the inability of detecting clean  electromagnetic signals from the depths of the central engine
and complexities involved in a first principles description of the underlying processes. 
Recently however, strong momentum
has been gained at the observational level~\cite{McClintock:2013vwa} (with further exciting
opportunities via near-future VLBI observations, e.g.~\cite{Broderick:2011mk}) as well as in the theoretical
front thanks to simulations of relevant systems, e.g.~\cite{Komissarov:2007rc,Palenzuela:2010nf,Neilsen:2010ax,2010ApJ...711...50T}). A large body of such 
simulations models the behavior
of the plasma and accompanying electromagnetic fields by adopting a force-free electrodynamics (FFE) 
approach~\cite{Goldreich:1969sb,1977MNRAS.179..433B}. Such model assumes (the physically realistic condition) that in the magnetosphere region, the matter contribution to the stress-energy tensor is negligible when compared to that of the the electromagnetic field. This assumption accounts for the plasma behavior implicitly through suitable
constraints, allowing one to derive a closed set of evolution equations that involve only the electric and magnetic fields, suitably coupled to a description of the spacetime curvature.

These equations constitute a highly non-linear hyperbolic PDE system as long as $F_{ab} F^{ab} = 2(B^2-E^2) \geq 0$ (i.e. the system is magnetically dominated) \cite{2011CQGra..28m4007P,Harald}, of which few analytical solutions are known \cite{Menon:2011zu,Michel1973,1976MNRAS.176..465B,
Lyutikov:2011tq,Brennan:2013jla,Brennan:2013ppa,Lupsasca:2014pfa,Yang:2014zva}. As a result, much of our current detailed
understanding has been obtained via numerical simulations which 
have provided important insights in the behavior of force-free, black hole systems. For instance,
how the black hole-plasma interaction sustains a steady and energetic Poynting flux as well
as the dependency of the latter with black hole spin~\cite{Contopoulos:1999ga,Komissarov:2002my,Spitkovsky:2006np,Kalapotharakos:2008zc,Palenzuela:2010xn, 
Neilsen:2010ax,Parfrey:2011ta,Asano:2005di,Cho:2004nn,Uzdensky:2004qu,
McKinney:2006sc,Timokhin:2006ur,Yu:2010bp,
Palenzuela:2010nf,Alic:2012df,Kalapotharakos:2011db,Petri:2012cs}. Despite the knowledge that can be gained
through simulations, it is certainly desirable to obtain 
analytical or semi-analytical solutions for their invaluable power to provide further clarity, 
allow for a broader generality and to provide additional guidance to the simulations' results.

Among relevant scenarios, the regime of rapidly spinning black holes\footnote{Such as possibly Cyg X-1 and GRS 1915+105 \cite{McClintock:2006xd,Gou:2011nq,McClintock:2013vwa}} 
is of particular importance due to the challenges they present to numerical simulations and the seemingly
more intricate phenomena allowed. For instance, subtle differences in the dependence of Poynting flux luminosity on the spin of highly spinning black holes have already been indicated by simulations~\cite{2010ApJ...711...50T}.
Additionally, it has been suggested that rapidly spinning black holes \cite{Yang:2013uba, Yang:2012pj, Yang:2012he} possess slowly-decaying quasinormal modes, which may reveal nonlinear instabilities if the mode-mode coupling is sufficiently strong \cite{Yang:2014tla}. Further interesting phenomena in the plasma can consequently arise and
a first step towards understanding it requires examining the plasma behavior on a fixed background.
Further reasons for
studying this regime are provided by the Kerr/CFT duality conjecture \cite{Bredberg:2011hp,Compere:2012jk} which
relates the NHEK to a suitable conformal fiel theory in 2+1 dimensions. 
Therefore, analytical solutions on the (near-) extremal Kerr black hole background are particularly interesting. 

In our pursuit to find such solutions, we are fortunate in that the NHEK metric \cite{Bardeen:1999px,Guica:2008mu} that describes the near horizon region of extremal holes possess an enhanced symmetry as compared to the generic Kerr metric. This allows one to concentrate on obtaining highly symmetric (i.e. more restricted) FFE solutions. Earlier attempts in this direction include Ref.~\cite{Li:2014bta} that found singular partial solutions near the poles or at large radius, and in particular Ref.~\cite{Lupsasca:2014pfa} that made explicit and sophisticated use of the symmetries to find a large family of exact solutions that are explicitly known everywhere (albeit not magnetically dominated). 

In this paper, we use an alternative (to Ref.~\cite{Lupsasca:2014pfa}) FFE solving framework to reduce the force-free equations to a single non-linear ordinary differential equation (ODE). Namely, we adopt the geometric language of Refs.~\cite{Gralla:2014yja,Carter1979,1997MNRAS.286..931U,
1997MNRAS.291..125U,1997PhRvE..56.2181U,1997PhRvE..56.2198U,
1998MNRAS.297..315U,Yang:2014zva} that simplifies the exploitation of symmetry considerations, and impose self-similarity under the conformal transformations. We essentially work under the $H$ representation described in Ref.~\cite{Lupsasca:2014pfa} instead of the $L$ representation utilized in that paper. The resulting family of solutions also differs from those found in Ref.~\cite{Lupsasca:2014pfa}, and include those that are  magnetically dominated. As we will
discuss, the final ODE has the peculiarity that at light surfaces, its character changes from second
to first order making it delicate to solve via standard methods.
 We instead develop a new procedure that circumvents this difficulty and apply it to generate two specific regular and globally magnetically dominated solutions. As the existence of light surfaces is generic, we expect this method to be widely applicable. In addition, our approach of imposing constraints to help reduce the problem to a single ODE is systematic, and should also prove useful in other scenarios. Specifically for the NHEK problem, aside from contributing a pair of particular solutions without physical or mathematical pathologies (which has not previously been found in literature), our discussion also lays down all the necessary tools for generating more interesting conformally self-similar solutions in future explorations. 

The paper is organized as follows. We begin by summarizing the background information such as the NHEK metric and the geometric FFE formalism in Sec.~\ref{sec:Background}, before moving on to impose the conformal self-similarity condition and obtain a final stream equation in Sec.~\ref{sec:StreamEq}. In Sec.~\ref{sec:Properties}, we analyze some predictable properties of the solutions to this stream equation, and in Sec.~\ref{sec:minimization}, we propose a minimization-based method to solve it. We then present two non-singular and globally magnetically dominated solutions in Sec.~\ref{sec:Solution}. 

\section{The NHEK spacetime and the FFE equations \label{sec:Background}}

\subsection{The NHEK metric}
To obtain the NHEK metric, we begin with the Kerr metric in Boyer-Lindquist (BL) coordinates (denoted by the hat
$\wedge$ symbol)
\bea
ds^2 &=& -e^{2\nu} d\hat{t}^2 + e^{2\Psi}(d\hat{\phi}-\omega d\hat{t})^2 \notag \\
&&+ \rho^2(\Delta^{-1} d\hat{r}^2+ d\hat{\theta}^2)\,,
\eea
where
\bea
\rho^2 &=& \hat{r}^2 + a^2 \cos^2\hat{\theta}, \quad \Delta = \hat{r}^2 - 2M \hat{r} + a^2, \notag \\ 
\omega &=& \frac{2M \hat{r} a}{\Delta \rho^2} e^{2\nu}, \quad 
e^{2\nu} = \frac{\Delta \rho^2}{(\hat{r}^2+a^2)^2-\Delta a^2 \sin^2\hat{\theta}}, \notag \\ 
e^{2\Psi} &=& \Delta \sin^2\hat{\theta} e^{-2\nu}\,,
\eea
and then carry out the transformation 
\bea \label{eq:BLvsNHEK}
\theta = \hat \theta, \quad R = \frac{\hat r - M}{2M^2\zeta}, \quad T = \zeta\hat t, \quad \phi = \hat \phi - \frac{\hat{t}}{2M}\,, 
\eea
into NHEK coordinates, 
before taking the limit $\zeta \rightarrow 0$ (not directly evaluating the Kerr metric at $\zeta =0$ which is indeterminate) that stretches the horizon region $\hat r \approx M$ out along the $R$ direction, while setting $a=M$. This gives us finally
\bea\label{eqnhek}
ds^2 &=& 2M^2 \Gamma(\theta) \left [  -R^2 dT^2+\frac{d R^2}{R^2}+d \theta^2 \right. \notag \\
&&\left. 
+\Lambda^2(\theta) (d \phi+ R dT)^2   \right ]\,,
\eea
where
\begin{equation}
\Gamma(\theta) = \frac{1+\cos^2\theta}{2}\,,\quad  \Lambda(\theta) = \frac{2 \sin\theta}{1+\cos^2\theta}\,.
\end{equation}
In NHEK coordinates, the value of $R=0$ corresponds to the horizon at $\hat{r}=M$, while any finite $R$ corresponds to a point infinitesimally away from the horizon in the Boyer-Lindquist coordinates~\cite{Bredberg:2011hp}. 

Besides time-independence and rotational-invariance, we note that this NHEK spacetime possesses an additional continuous symmetry, namely the conformal symmetry
\begin{equation} \label{eq:ConfTrans}
R \rightarrow \lambda R,\quad T \rightarrow T/\lambda\,,
\end{equation}
with the symmetry generator 
\begin{equation}
\mathcal{H}_C = T \partial_T - R\partial_R\,.
\end{equation}
This symmetry corresponds to a rescaling of the $\zeta$ parameter in Eq.~\eqref{eq:BLvsNHEK} that does not affect the final metric. 

Following the discussion in Ref.~\cite{Gralla:2014yja}, we divide the total NHEK spacetime into a ``poloidal" subspace and a ``toroidal" subspace, with their respective area two-forms being
\begin{equation} \label{eq:Area1}
\epsilon^T =\sqrt{-g^T} dT \wedge d\phi,\quad \epsilon^P = \sqrt{g^P} dR \wedge d\theta\,,
\end{equation}
with, 
\begin{equation} \label{eq:Area2}
\sqrt{-g^T} = 2M^2 \Gamma(\theta) R \Lambda(\theta),\quad \sqrt{g^P} = 2M^2 \Gamma(\theta)\frac{1}{R} \,.
\end{equation}

\subsection{Non-extremal black holes}
Recall that 
the highest spin of astrophysical black holes has been estimated at 0.998 using thin-disk models \cite{Thorne:1974}.
It is then important to consider how to map a NHEK solution out to sub-extremal black-hole spacetimes. Fortunately it is possible to do so by writing down a transformation between the Boyer-Linquist (BL) coordinates of sub-extremal black holes to the NHEK coordinates.

Following the discussion in \cite{Hadar:2014hps}, we define $\kappa \equiv \sqrt{1-a^2} $ and the following coordinate system
\begin{equation}
r= \frac{\hat r - r_+}{r_+},\quad t = \frac{\hat t}{2 M}, \quad \phi = \hat \phi -\frac{\hat t}{2 M}\,,
\end{equation}
which is a simple transformation from the BL coordinates in Kerr. Here $r_+ \equiv M(1+\kappa)$ is the radius of the outer horizon. By taking the near horizon limit $r \ll 1$ in such coordinates, we find that the metric reduces to
\begin{align}\label{eqmnek}
ds^2 = & 2M^2 \Gamma(\theta) \left [ -r(r+2 \kappa) dt^2+\frac{d r^2}{r(r+2 \kappa)} \right . \nonumber \\
 + & \left . d\theta^2+\Lambda^2(\theta)(d\phi+(r+\kappa)dt)^2\right ]\,.
 \end{align}

More importantly, the above metric can be transformed to a NHEK metric by the following transformation
\begin{align}\label{eqnheknek}
T = & -e^{- \kappa t} \frac{r+\kappa}{\sqrt{r(r+2\kappa)}}\,,\nonumber \\
R = & \,\frac{e^{\kappa t}}{\kappa} \sqrt{r(r+2\kappa)}\,, \nonumber \\
\Phi = & \,\phi -\frac{1}{2} \log \frac{r}{r+2\kappa}\,, 
\end{align}
which justifies applying the NHEK solutions to the case of sub-extremal black holes. In addition, under the same transformation, it is straightforward to show that the conformal Killing vector in NHEK maps to the Killing vector in the time direction of the BL coordinates
\begin{equation}
\mathcal H_C \leftrightarrow -\frac{1}{\kappa}\partial_{t}\,.
\end{equation}
In later sections, we obtain FFE solutions without exact conformal symmetries in the NHEK coordinates, and the same solutions describe time-dependent (in BL coordinates) FFE solutions of sub-extremal black holes.

\subsection{Force-free equations}

Let us turn now to the force-free equations in NHEK.
The assumption that the plasma contribution to the stress-energy tensor is negligible implies~\cite{Parfrey:2011ta,Palenzuela:2010xn} 
\bea \label{eq:FFCond}
0=\nabla_a T^{ab}\approx \nabla_a T_{\text{EM}}^{ab} =-F^{ab}j_b\,, 
\eea
which is called the force-free condition, as its spatial part implies the vanishing of the Lorenz force on the plasma. It has long been known that the Maxwell equations can be written in their most economic form using differential forms, in which case they become
\begin{equation} \label{eq:Maxwell}
d F=0\,, \quad d{}^*F =J\,, 
\end{equation}
where $F$ is the Faraday tensor, while $J$ is the current three form - the Hodge dual to the 4-D current one form $j$. It has also been shown that the force-free condition \eqref{eq:FFCond} can be written in the same geometric language. In particular, the force-free condition implies (but not necessarily vice versa) that the field must be degenerate: $F \wedge F =0$, and that $F$ can be written as the wedge product of two $1$-forms:
\begin{equation}
F = d \phi_1 \wedge d \phi_2\,,
\end{equation}
where $\phi_1$ and $\phi_2$ are called Euler potentials. 
In the case that the background metric and the FFE solution are both stationary and axis-symmetric, Refs.~\cite{1997PhRvE..56.2198U,Gralla:2014yja} further showed that $\phi_1$ and $\phi_2$ can be written as
\begin{equation} \label{eq:EulerPot}
\phi_1 = \psi(R, \theta),\,\quad  \phi_2 = \psi_2(R,\theta)+\phi-\Omega_F (\psi) T\,,
\end{equation}
where $\psi$ is the magnetic flux function, in the sense that 
\begin{equation} \label{eq:fluxdef}
\psi(R,\theta) = \frac{1}{2\pi} \int_{\mathcal P} F\,,
\end{equation}
with $\mathcal{P}$ being any two dimensional surface bounded by a loop of constant $(T,R,\theta)$ but varying $\phi$. It is also frequently referred to as the stream function. 

The full force-free condition then translates into
\begin{equation} \label{eq:FFCondition}
d \phi_1 \wedge J=0=d \phi_2 \wedge J\,.
\end{equation}
Using Eq.~\eqref{eq:Maxwell}, these two expressions can be rewritten as 
\bea \label{eq:FFEqns}
d \phi_1 \wedge d{}^* F= 0\,, \quad d \phi_2 \wedge d{}^* F =0\,,
\eea
where the first expression corresponds to the conservation of energy and angular momentum, while the second is called the stream equation \cite{Gralla:2014yja}. 

Finally, it is important to note that there is a hidden constraint for FFE that is not automatically guaranteed by Eqs.~\eqref{eq:FFEqns}. Namely the solution must be magnetically dominated with $B^2-E^2 \geq 0$. The physical significance of this condition can be understood by noting that $(E\times B)/B^2$ is the drift velocity for the advection of the charge density \cite{Spitkovsky:2006np,Parfrey:2011ta}. The inequality $E^2 > B^2$ then implies superluminal motion for the plasma. A symptom of this unphysical scenario is that some characteristic speeds of the force-free equations become complex, so the evolution system ceases to be 
hyperbolic~\cite{2011CQGra..28m4007P,Harald}. The Eqs.~\eqref{eq:FFEqns} do not enforce this condition however, as they are derived without referencing the plasma equations of motion, and simply do not know that superluminal plasma motion is an issue. Therefore, magnetic dominance should be checked after solving Eqs.~\eqref{eq:FFEqns}.

\section{The stream equation for self-similar solutions \label{sec:StreamEq}}
In general, the solutions to the FFE equations can be less symmetric than the underlying spacetime metric. 
However, imposing extra symmetries can help us narrow down the choice for $\Omega_F$, $\psi_2$ and $\psi$, at the cost of restricting ourselves to a more specialized subset of solutions. From here on, we will consider solutions that are self-similar under the conformal transformation \eqref{eq:ConfTrans}, namely that the Faraday tensor transforms into some constant times itself (we also require time-stationary and rotational symmetry, so that we can use expression \ref{eq:EulerPot}). 
Furthermore, because we are trying to constrain and simplify the FFE equations as much as possible, we further demand that the two Euler potentials be separately self-similar (therefore in general, our solution is a special subset of all conformally self-similar solutions). 
We have explicitly 
\bea \label{eq:Faraday}
F = d\psi(R,\theta) \wedge \left(d\psi_2(R,\theta) + d\phi - \Omega_F(\psi) dT\right)\,,
\eea
and as $d\phi$ is invariant under Eq.~\eqref{eq:ConfTrans}, we need $\Omega_F dT$ and $d\psi_2$ to also be invariant, which is easily accomplished with $\Omega_F =g(\theta) R$ and $\psi_2 =h(\theta)$. We also want $d\psi$ to be self-similar, therefore $\psi$ should have a dependence on $R$ of the form $\psi(R,\theta) = R^\alpha f(\theta)$, with the power $\alpha$ yet to be determined. We note that this means $\psi=0$ on the horizon for any $\alpha>0$. Indeed, the condition that 
\bea
d\psi = \alpha R^{\alpha-1}f(\theta)dR + R^{\alpha} f'(\theta) d\theta
\eea
and subsequently $F$ as given by Eq.~\eqref{eq:Faraday} remain regular at the horizon $R=0$ requires $\alpha \geq 1$ (of all the coordinate one forms, only $dT$ diverges as $R^{-1}$ on the horizon, and this cancels with the $R$ factor in $\Omega_F$ within Eq.~\eqref{eq:Faraday}, so $F$ is regular as a whole as long as $d\psi$ is regular). 
In addition, since $\Omega_F$ is a function of $\psi$ only, the function $g(\theta)$ can be expressed in terms of $f(\theta)$ as
\begin{equation}
g(\theta) = C f(\theta)^{1/\alpha}\,,
\end{equation}
with $C$ being some constant, and $\Omega_F =C \psi^{1/\alpha}$. We notice that the self-similarity property of the above solution can be expressed in terms of the Lie-derivative
\begin{equation}
\mathcal L_{\mathcal H_C} F =-\alpha F\,.
\end{equation}
Notice that for any $\alpha \neq 0$ the associated solutions do not respect the exact conformal symmetry, so that they are time-dependent when mapped to the BL coordinates of near-extremal Kerr black holes.

We can also compute the polar current $I$, which is defined as \cite{Gralla:2014yja} 
\begin{equation}\label{eqdefi}
* (d\psi \wedge d \psi_2)  = \frac{I }{2\pi} d T \wedge d\phi\,,
\end{equation} 
and can be seen as essentially a substitute for $\psi_2$ or $h$. Explicitly, we find that 
\begin{align}
* (d\psi \wedge d \psi_2) = &  *\left ( \alpha \frac{\psi}{R} h'(\theta) dR \wedge d \theta\right ) \nonumber \\
= & \alpha \,\psi\, h'(\theta) R \Lambda(\theta) dT \wedge d \phi\,,
\end{align}
where we have used $*\epsilon^P=\epsilon^T$, as well as Eqs.~\eqref{eq:Area1} and \eqref{eq:Area2}. 
We have then
\begin{equation}\label{eqic1}
I = 2\pi\, \alpha  \, h'(\theta)\, \Lambda(\theta) \, \psi R\,, 
\end{equation}
where the prime denotes derivative against $\theta$. 

By applying the energy and angular momentum conservation (i.e. the first equation in \ref{eq:FFEqns}), one concludes that $I=I(\psi)$, which implies $d \psi \wedge dI =0$ (See Eq.~(75) and (76) of \cite{Gralla:2014yja} for more details). For our specific case, this means that 
\bea
\left ( \frac{\alpha f(\theta)}{R} dR+f'(\theta) d\theta\right ) \wedge f(\theta)\left[h'(\theta)\Lambda(\theta)dR \frac{}{}\right. \notag \\
\left. \frac{}{}+(h' \Lambda)' R d\theta\right]=0\,.
\eea
Assuming $f(\theta) \neq 0$ as well as $h' \neq 0$ to avoid trivial solutions, we then must have 
\begin{align}
\frac{f'}{\alpha f} - \frac{(\Lambda h')'}{\Lambda h'} =0\,,
\end{align}
which further implies
\begin{equation}
h' \Lambda =  D f(\theta)^{1/\alpha}\,,
\end{equation}
with $D$ being a constant. The current is then 
\begin{equation} \label{eq:Current}
I =2\pi \, \alpha  \, D \psi^{1+1/\alpha}\,.
\end{equation}
By requiring the solution to be conformally self-similar, we have thus fixed the functional forms of both $\Omega_F$ and $I$, which is one of the toughest hurdles to obtaining analytical FFE solutions \cite{Gralla:2014yja}.

We have now only the stream equation --the second equation in (\ref{eq:FFEqns})-- that still needs to be satisfied. Expressed under the quantities appearing in the Euler potentials, the stream equation takes the form \cite{Gralla:2014yja}
\begin{equation}
\nabla_a (|\eta |^2 \nabla^a \psi)+\Omega_{F\,,\psi} \langle dt, \eta \rangle |d \psi|^2-\frac{I \, I_{,\psi}}{4 \pi^2 g^T}=0\,,
\end{equation}
where $||$ and $\langle \rangle$ simply denote contractions using the NHEK metric, and 
\bea
\eta \equiv d\phi - \Omega_F(\psi)dT\,.
\eea
The quantities $|\eta|^2$ and $\langle dt, \eta \rangle$ are given in Eqs. (87)-(89) of Ref.~\cite{Gralla:2014yja}, which for our case become
\begin{align}
|\eta|^2 &= \frac{1}{2 M^2 \Gamma(\theta)}\left [ \frac{1}{\Lambda^2(\theta)}-\left ( \frac{\Omega_F}{R}+1\right )^2\right ] \nonumber \\
&= \frac{1}{2 M^2 \Gamma(\theta)}\left [ \frac{1}{\Lambda^2(\theta)}-\left ( g(\theta)+1\right )^2\right ]\,,\label{eq:eta2}\\
\langle dt, \eta \rangle &=\frac{1}{2 M^2 \Gamma(\theta) R}[g(\theta)+1]\,,
\end{align}
and so the terms in the stream equation are
\begin{align}
-\frac{I \, I_{, \psi}}{4 \pi^2 g^T} = & \frac{\alpha^2 D^2 (g/C)^{\alpha+2}R^{\alpha} (1+1/\alpha)}{4 M^4 \Gamma^2 \Lambda^2} \,,\nonumber \\
\Omega_{F\,, \psi} \langle dt, \eta \rangle |d \psi|^2 = & \frac{C^{-\alpha} \alpha R^{\alpha} g^{\alpha-1} (1+g)[(g')^2+g^2]}{4 M^4 \Gamma^2} \, \label{eq:SE1}
\end{align}
and
\begin{widetext}
\bea\label{eq:SE2}
\nabla_a (|\eta |^2 \nabla^a \psi)&=& 
\frac{R^{\alpha} \alpha (g/C)^{\alpha}}{4 M^4 g^2 \Gamma^2}\left\{
\left(\frac{}{}g''+(1+\alpha)g \right)g\left(\frac{1}{\Lambda^2}-(1+g)^2 \right)
-\left(\frac{1-\alpha}{\Lambda^2}+(1+g)\left(\frac{}{}\alpha-1 +(1+\alpha)g\right) \right)(g')^2 
\right. \nonumber \\ && \left.
-\left(\frac{}{}\cot\theta+\cos\theta \Lambda\right)\left(\frac{1}{\Lambda^2}+(1+g)^2 \right)gg'
\right\} \,.
\eea
\end{widetext}
Combining the three expressions and multiplying by $C^{\alpha}$, we can then replace $D^2/C^2$ with a parameter $\xi$, and note that the remaining $C$ can be factored out of the equation and ignored. We can also divide out the $R^{\alpha}$ term so the equation reduces to one depending on $\theta$ only. We can further multiply the equation by $\Lambda^2 \Gamma^5$ to ensure all terms in it remain regular at the poles ($\theta=0$ or $\pi$) as well as to get rid of spurious overall factors. It is this form of the equation that we will solve later in Sec.~\ref{sec:Solution}. For reference, we note that when specializing to $\alpha=1$ and defining $x=\cos\theta$, we can rewrite the stream equation to the simplified form of
\bea \label{eq:StreamEq}
&&\left [ \frac{\Gamma^2}{1-x^2}-(1+g)^2\right ] \left (g_{,xx}-\frac{2 x}{1-x^2} g_{,x} +\frac{2 g}{1-x^2} \right ) \notag \\
&&+\left [ \frac{\Gamma}{1-x^2}-\frac{(1+g)^2}{\Gamma}\right ]_{, x} g_{,x} \Gamma \notag \\
&&+(1+g)\left [(g_{,x})^2+\frac{g^2}{1-x^2} \right ]+ \frac{2\xi g^3 \Gamma^2}{(1-x^2)^2}=0\,.
\eea
For the rest of this paper, we will frequently use $\alpha=1$ as a concrete example, but our discussions easily generalize to $\alpha > 1$. 

We stress that the change of variable into $x$ is more than a notational convenience. The regularity of $g$ at $\theta=0$ or $\pi$ requires that $g'|_{\theta=0,\pi}=0$ (otherwise $g$ as an axisymmetric scalar field will not have a well-defined first derivative at the poles). Since 
\bea
g'=-\sqrt{1-x^2} g_{,x}\,,
\eea
the regularity condition for radial magnetic field simply translates into $g_{,x}$ not diverging at $x=\pm 1$. There is another physical boundary condition on the poles that must be taken into account. Recall that the loop on the rim of $\mathcal{P}$ in Eq.~\eqref{eq:fluxdef} shrinks to a single point at $\theta=0$ or $\pi$, so we can choose $\mathcal{P}$ with vanishing area. Therefore, the magnetic flux across it should vanish if $F$ does not diverge there. In other words, we need $g|_{\theta=0,\pi}=0$. We further note that the stream equation is symmetric under a $\theta \rightarrow \pi-\theta$ reflection, and so we can obtain a reflection symmetric solution if our boundary conditions respect this symmetry. We thus impose $g'|_{\theta=\pi/2}=0$, and concentrate only in the region $0\leq \theta \leq \pi/2$, with the understanding that the other half of the solution can be obtained by symmetry. Note that this last condition can be changed if one desires non-reflection-symmetric solutions. 

\section{Properties of small magnitude solutions \label{sec:Properties}}
Even though the stream equation is highly non-linear and non-trivial, we can still predict some properties of its solutions under special circumstances. Such a situation arises when $g$ has a small magnitude, so that many of the non-linear terms in the equation become negligible.
This results in great simplification with respect to the treatment of the light surfaces, defined by the condition $|\eta|^2=0$ that leads to the coefficient of the $g''$ term vanishing and the equation locally reducing to first order. When the nonlinear terms are negligible, one can in fact predict the locations of the light surfaces by ignoring $g$ in Eq.~\eqref{eq:eta2}, and solving $|\eta|^2=0$ to find that a light surface lies at $x=\sqrt{-3+2\sqrt{3}}$. Furthermore, we can also predict that $B^2-E^2$ should change sign at the light surface.

To compute $B^2-E^2$, we first note that the Faraday tensor is given by 
\begin{widetext}
\bea \label{eq:FaradaySol}
F = \alpha\frac{1}{C}\left(\frac{Rg}{C}\right)^{\alpha-1} \left( g(\theta) dR + R g'(\theta) d\theta \right)\wedge \left(\sqrt{\xi} \frac{g(\theta)}{\Lambda(\theta)}d\theta + d\phi - Rg(\theta)dT \right)\,,
\eea
which when setting $M=1$ for simplicity and specializing to $\alpha=1$ for concreteness (generic $\alpha$ expressions can be recovered by simply replacing $1/C$ in the equations below by $(\alpha/C) (Rg/C)^{\alpha-1}$), gives 
\bea 
E^a &=& F^{ab}T_b = \frac{1}{C}\left( 0,0, -\frac{g(1+g)R^2 \sqrt{\Gamma}}{2\sqrt{2}\Gamma^2}, -\frac{(1+g)g' R \sqrt{\Gamma}}{2\sqrt{2}\Gamma^2}\right)
\,, \label{eq:EField}\\
B^d &=& \frac{1}{2} \epsilon^{abcd}F_{ab} T_c = \frac{1}{C}\left( 0, -\frac{g^2\sqrt{\xi} R\sqrt{\Gamma}}{2\sqrt{2}\Gamma^2 \Lambda^2},
-\frac{g' R^2 \sqrt{\Gamma}}{2\sqrt{2} \Gamma^2 \Lambda}, 
\frac{g R \sqrt{\Gamma}}{2\sqrt{2} \Gamma^2 \Lambda}
\right)\,, \label{eq:BField}
\eea
and subsequently
\bea 
\label{eq:Poynting}
P^a = \frac{1}{C^2}\left(0,-\frac{(1+g)(g^2+ g'^2) R^2}{4 \sqrt{2} \Gamma^{5/2} \Lambda^2 }, 
\frac{(1+g)g^2 g'R^3 \sqrt{\xi}}{4 \sqrt{2} \Gamma^{5/2} \Lambda},
-\frac{(1+g)g^3  R^2 \sqrt{\xi}}{4 \sqrt{2} \Gamma^{5/2} \Lambda}
\right)\,,
\eea
\end{widetext}
with
\bea
T_a = (-\sqrt{2}R\sqrt{\Gamma(\theta)},0,0,0)
\eea
being the one form normal to the $T=const$ spatial slices. 
All the vectors are in the coordinate basis $(\partial_t,\partial_{\phi},\partial_r,\partial_{\theta})$ and similarly for the one forms.
Note that $C$ is nothing more than a scaling factor for $F$, and we will set it to $1/2$ from here on. We also draw attention to the appearance of $\xi$ in the coefficient of the $\partial_{\phi}$ component of $B$ and nowhere else. This provides a physical significance for $\xi$ as generating the spiralling of the $B$ field lines in the longitudinal direction. In particular, $P^a$ is purely in the $\partial_{\phi}$ direction when $\xi=0$. Furthermore, it is easy to verify that the force-free constraint of $E\cdot B=0$ (a consequence of Eq.~\ref{eq:FFCond}) is indeed satisfied.  
 
From Eqs.~\eqref{eq:EField} and \eqref{eq:BField}, it is then straightforward to show that 
\bea \label{eq:BSqrMinusESqr}
B^2 - E^2 &=& \frac{R^2}{\Gamma^2\Lambda^2}\left[ g^4 \xi \right. \notag \\ &&\left. 
+ (g^2 +g'^2)(1-(1+g)^2\Lambda^2) \right],
\eea  
(note $R$ factors out and does not affect the sign of this expression),
which is in fact the explicit form for the more generic expression (note the correction as compared to Eq.~(66) of Ref.~\cite{Gralla:2014yja})
\bea \label{eq:MagDom2}
B^2-E^2= \frac{I^2}{4\pi^2(-g^T)} + |d\psi|^2 |\eta|^2\,. 
\eea
Substituting in Eq.~\eqref{eq:Current}, we see that $I^2 \propto g^4$ when $\alpha=1$ and can be ignored when $|g|$ is small. On the other hand, $|d\psi|^2$ is always positive as $d\psi$ is spacelike, while $\eta$ changes character at the light surface from space-like to time-like, so $|\eta|^2$ and subsequently $B^2-E^2$ changes sign at the light-surface. Indeed, we can explicitly substitute the expression for $g$ (by solving $|\eta|^2=0$) and $g'$ (by solving the locally first order stream equation) at the light surface into Eq.~\eqref{eq:BSqrMinusESqr} and verify that $B^2-E^2=0$ at $x=\sqrt{-3+2\sqrt{3}}$. 

Finally, in order to find a physically realistic solution that is globally magnetically dominated, we note that the first term in Eq.~\eqref{eq:MagDom2} is always positive, so a large $g$ magnitude is expected to benefit our task (although the magnitude of the second term may also increase). 

\section{Solving the stream equation with a residual minimization method \label{sec:minimization}}
A large $g$ magnitude significantly complicates the solution finding process. 
Traditionally, one can solve the stream equation separately on two sides of a fixed light surface and attempt to match them across the light surface as smoothly as possible by varying $\Omega_F$ and $I$ as functions of $\psi$ \cite{Contopoulos:1999ga,Uzdensky:2004qu,Timokhin:2005mx,Gruzinov:2006nn,Contopoulos:2012py,Nathanail:2014aua}. 
In our case, the one dimensional ODEs such as Eq.~\eqref{eq:StreamEq} can also contain light surfaces, but their locations are not known a priori when $|g|$ is large, because they depend on $g$, and as we have already fixed $\Omega_F$ and $I$, the condition of smooth matching 
should instead exert itself through fixing the locations (and the number) of the light surfaces. 
For the rest of this paper, we concentrate on finding solutions that are as smooth across the light surface as possible \footnote{Note we do not necessarily require $C^{\infty}$ where there is a light surface, but we do however prefer the solution to be at least $C^{2}$ 
across the light surface. Otherwise $g'$ and/or $g''$ will not be well-defined there and we will only have a weak solution. Physically, it is also reasonable to expect that smoother solutions would require less dramatic non-FFE physics to be present at the light surfaces. \label{ft:1}}, but we note that if one can live with more singular behaviors, then the family of admissible solutions is much larger. 

\begin{figure}[t]
  \centering
  \begin{overpic}[width=0.95\columnwidth]{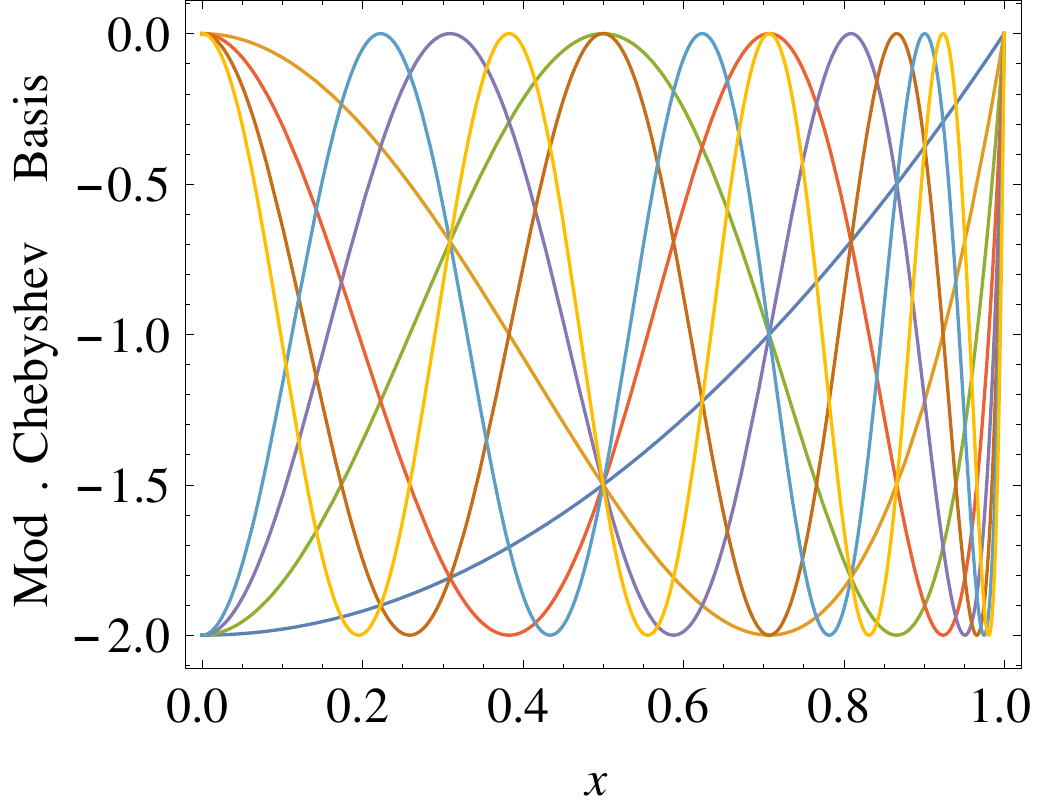}
\end{overpic}
  \caption{The modified Chebyshev polynomials used for the basis decomposition of $g$. }
	\label{fig:BasisPoly}
\end{figure}

At first sight, the most straightforward way to solve the equations is through a shooting method. Here one
imposes two boundary conditions ($g|_{x=1}=0$ and a value for $g'|_{x=1}$ that must be adjusted) at $x=1$ and marches the solution\footnote{For instance, using a readily available numerical ODE solving routine from popular software packages} towards $x=0$. Simultaneously, one imposes a pair of boundary conditions ($g'|_{x=0}=0$ and an
adjustable value for $g|_{x=0}$) at $x=0$ and marches the solution towards $x=1$. One then adjusts $g'|_{x=1}$ and $g|_{x=0}$ so that the two solutions intersect at a single light surface and match relatively smoothly across it. The problem with this strategy is that generically, $g''$ would diverge near the light surfaces in order to stay relevant (because its coefficient vanishes there) and be able to contribute to the balancing of the stream equation. Subsequently, $g'$ and $g$ usually also diverge, unless one has educated guesses so that the choice of $g'|_{x=1}$ and $g|_{x=0}$ matches the ``correct'' smooth solution that does not need a non-vanishing $g''$ term to balance the equation near the light surfaces. 

Alternatively, 
one may try to address this issue by adopting a different strategy of fixing the location of the light surface first, and solve for the $g$ value that satisfies $|\eta|^2=0$, as well as the $g'$ value that satisfies the locally first order stream equation. Using these as boundary conditions at the light surface, one then marches the solutions towards $x=0$ and $x=1$ (i.e. in the reverse direction of the previous strategy) while varying the light surface location to try and match the boundary conditions there. The difficulty here is that one can not impose two boundary conditions at places where the stream equation is first order. One can nevertheless impose them at locations straddling the light surface but slightly off of it, say by $\pm \delta$. However, the solutions generically show a sensitive dependence on $\delta$, essentially because one still finds the diverging solutions. Here, the solutions diverge at the ``right'' rate such that $g$ and $g'$ values at the offset locations are as specified. So once again, this strategy is only useful for finding non-singular solutions when a good initial guess is provided as to what the smooth solution should be. 

From the discussion above, it is clear that
the essence of the problem one faces is that with these strategies one is restricting to the space of \emph{exact} (ignoring numerical error) weak solutions to the stream equation, which is mostly populated by diverging solutions that make it difficult to single out the non-diverging (and possibly strong) ones. This problem is further
complicated at the numerical level where numerical errors make it difficult to latch onto the physical
solution exactly. This suggests a different strategy: to work within the space of \emph{approximate} but non-divergent solutions --the solution we look for is in the intersection of these two spaces-- and develop 
a method to consistently approach the physical solution. To this end, we decompose $g$ into a functional basis $\{f_i\}$ satisfying 
\bea \label{eq:BasisBC}
f_i'|_{x=0}=0 =f_i|_{x=1}\,, \quad \forall i \,, 
\eea
so that the boundary conditions for $g$ are automatically satisfied when we include a finite number of basis functions. We then alter the coefficients of decomposition using some minimization routine in order to minimize the residual $\mathcal{L} g$ (where $\mathcal{L}$ is a differential operator such that the stream equation is $\mathcal{L}g=0$). 
The advantage of this method is that we only ever apply $\mathcal{L}$ in its natural ``forward'' direction, never needing to compute its inverse or the inversion of any of its components. Therefore, we do not encounter any problem even when $\mathcal{L}$ becomes degenerate, and we do not need any prior knowledge or expectation on where the light surfaces would be, or even how many there are.

One set of functional basis that satisfies Eq.~\eqref{eq:BasisBC} can be obtained by taking away a constant $1$ from the Chebyshev polynomials 
\footnote{Note that other basis functions such as sinusoidal functions are also possible candidates.}
of the first kind and of even orders. We shall refer to them as the modified Chebyshev basis, and the first few of them are plotted in Fig.~\ref{fig:BasisPoly}. We note that including the polynomials of odd orders will preserve the $f_i|_{x=1}=0$ condition, but relax the $f'_i|_{x=0}=0$ condition. Once we decompose $g$ into these functional basis, we can utilize minimization algorithms such as a simple selective (reject or accept a step depending on whether it makes an improvement) random walk in the expansion coefficients space to minimize the normalized $L_2$ norm of $\mathcal{L}g$, i.e. $\int_0^1 dx (\mathcal{L}g)^2/\int_0^1 dx g^2$. 

We note here that the even order Chebyshev polynomials form a complete basis for even functions ($f(x)=f(-x)$) in the interval $[-1,1]$. Also, when $f(1)=0$ the same set of coefficients are valid for both decompositions into Chebyshev and modified Chebyshev polynomials, thus the modified Chebyshev polynomials also form a complete basis for the functions satisfying our desired boundary conditions. 
Therefore, if we include enough number of basis functions, we can in principal approximate the higher-order non-smooth behaviour of $g$ at the light surfaces, or even possibly the diverging solutions. In practice, as we only supply a few basis functions, our trial function is relatively smooth and thus is better able to approximate the smoother solutions. Therefore, we expect this residual minimization method to preferentially home-in on the smoothest solution possible, which is in fact a desired property (see footnote \ref{ft:1}). 
We note however, this method can be numerically expensive, especially if one utilizes the selective random walk procedure without any optimization. Therefore we combine it with the traditional techniques discussed earlier, using the residual minimization routine to provide educated initial guesses for the ODE integration. 

\section{Globally magnetically dominated solutions \label{sec:Solution}}
\begin{figure*}[t,b]
    \begin{overpic}[width=0.95\columnwidth]{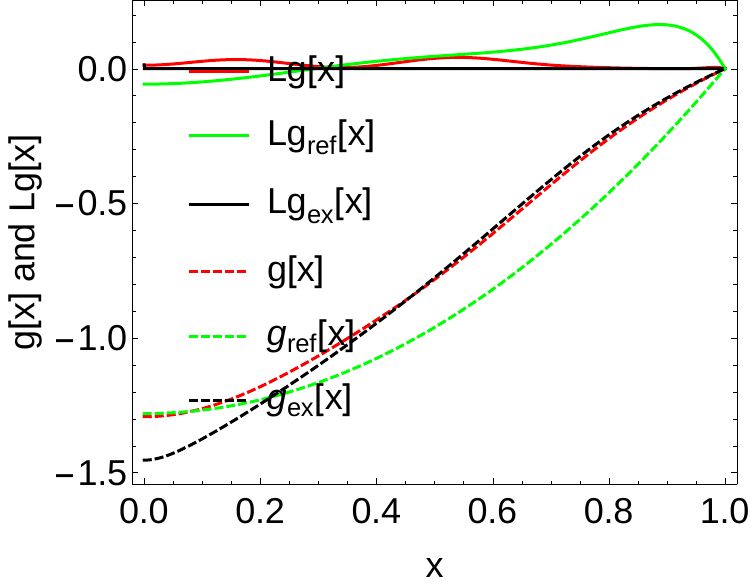}
\put(90,2){(a)}
\end{overpic}
   \begin{overpic}[width=0.95\columnwidth]{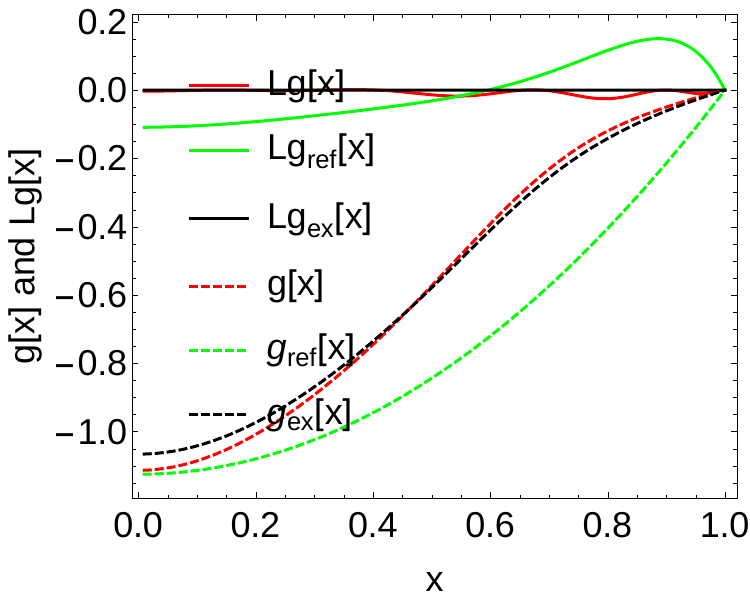}
\put(90,2){(b)}
\end{overpic}
  \caption{(a): An example solution with large magnitude for $g$ at $\xi =0$ and computed with $7$ modified Chebyshev basis functions. The renormalized initial guess $g_{\text{ref}}$, the outcome of selective random walk $g$ and the final solution $g_{\text{ex}}$, as well as their residuals (dashed lines) 
$\mathcal{L}g_{\text{ref}}$, $\mathcal{L}g$ and $\mathcal{L}g_{\text{ex}}$ are plotted.  (b): The same plot for another solution with $\xi=1$. 
}
	\label{fig:SolLargeg}
\end{figure*}

\begin{figure*}[t,b]
  \begin{overpic}[width=0.95\columnwidth]{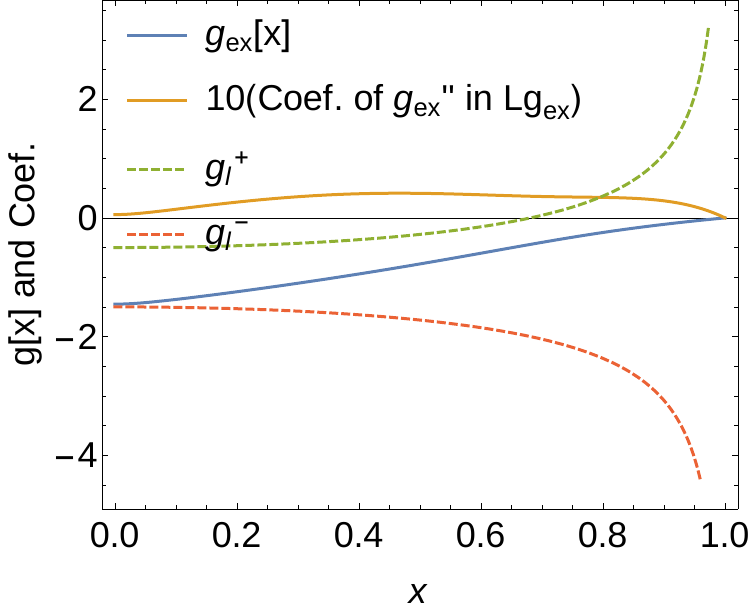}
\put(90,2){(a)}
\end{overpic}
  \begin{overpic}[width=0.95\columnwidth]{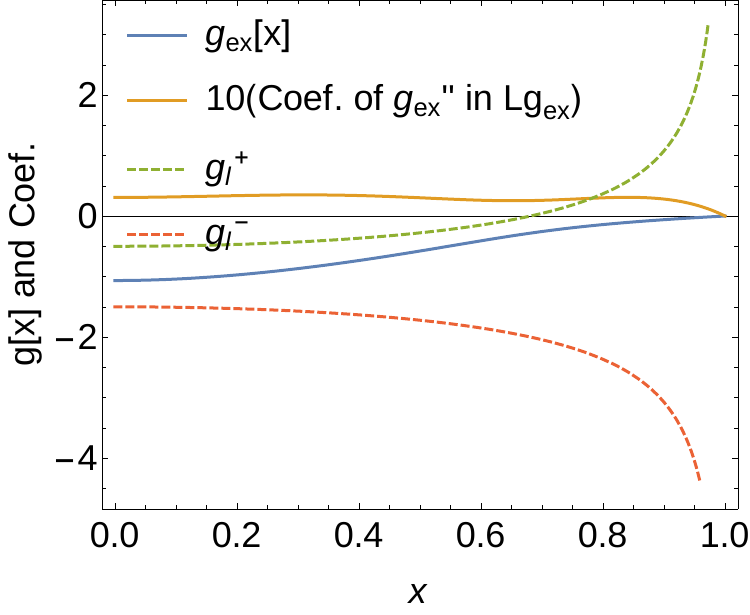}
\put(90,2){(b)}
\end{overpic}
  \caption{ 
(a): For the $\xi=0$ solution. Ten times the coefficient of the $g''_{\text{ex}}$ term in $\mathcal{L}g_{\text{ex}}$ vanishes at $x=1$ and nearly vanishes at $x=0$, but $g_{\text{ex}}$ only nearly intersects $g_l^{\pm}$ at $x=0$. We have also shown as a thin black line, the location of zero as a reference. 
(b): Same as (a) but for the $\xi=1$ solution. For this solution, it is more clear that the the equatorial plane is not a light surface.}
	\label{fig:SolLargegLS}
\end{figure*}

\begin{figure*}[t,b]
  \begin{overpic}[width=0.95\columnwidth]{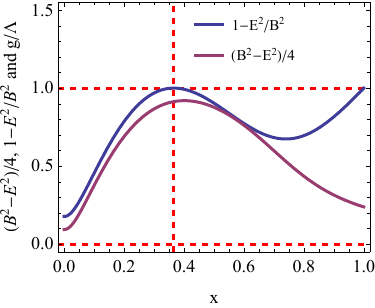}
\put(90,2){(a)}
\end{overpic}
  \begin{overpic}[width=0.95\columnwidth]{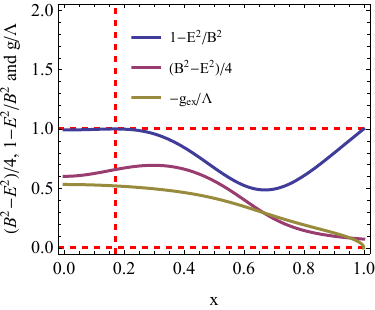}
\put(90,2){(b)}
\end{overpic}
  \caption{ 
(a): For the $\xi=0$ solution. The $(B^2-E^2)/4$ (to fit into same figure) and normalized $1-E^2/B^2$ values for $g_{\text{ex}}$ at $R=1$ (the sign of these quantities are $R$ independent, see Eq.~\ref{eq:BSqrMinusESqr}). The solution is globally magnetically dominated as the curves never fall below the horizontal dashed red line at $0$. We have also shown a horizontal line at $1$ and a vertical line at a location where $1-E^2/B^2=1$. We have $|E|=0$ at this location. 
(b): Same as (a) but for the $\xi=1$ solution, showing that this solution is also globally magnetically dominated. We have also plotted $g_{\text{ex}}/\Lambda$ to show that this quantity does not diverge at $x=1$ even though $\Lambda \rightarrow 0$ there.
}
	\label{fig:SolLargegMagDom}
\end{figure*}

\begin{figure*}[t,b]
    \begin{overpic}[width=0.32\textwidth]{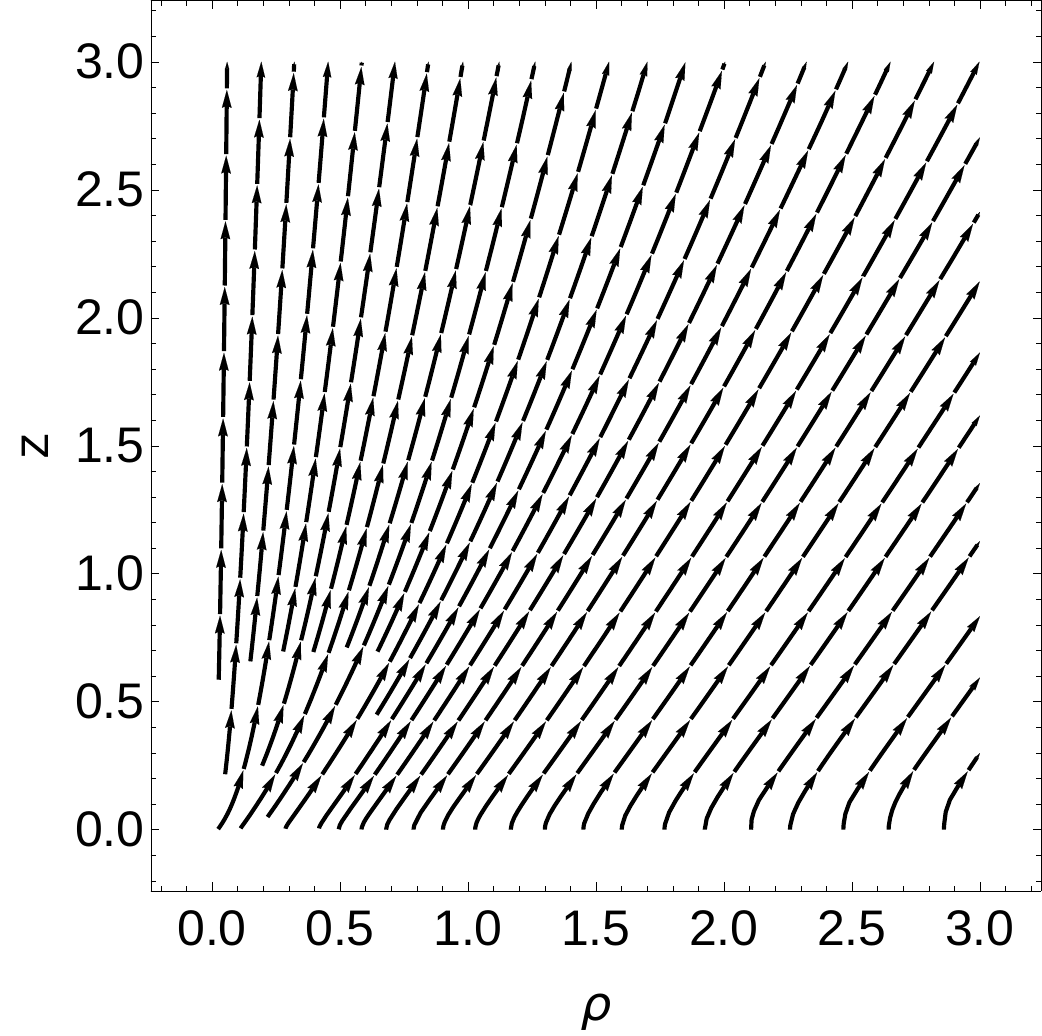}
\put(90,2){(a)}
\end{overpic}
  \begin{overpic}[width=0.32\textwidth]{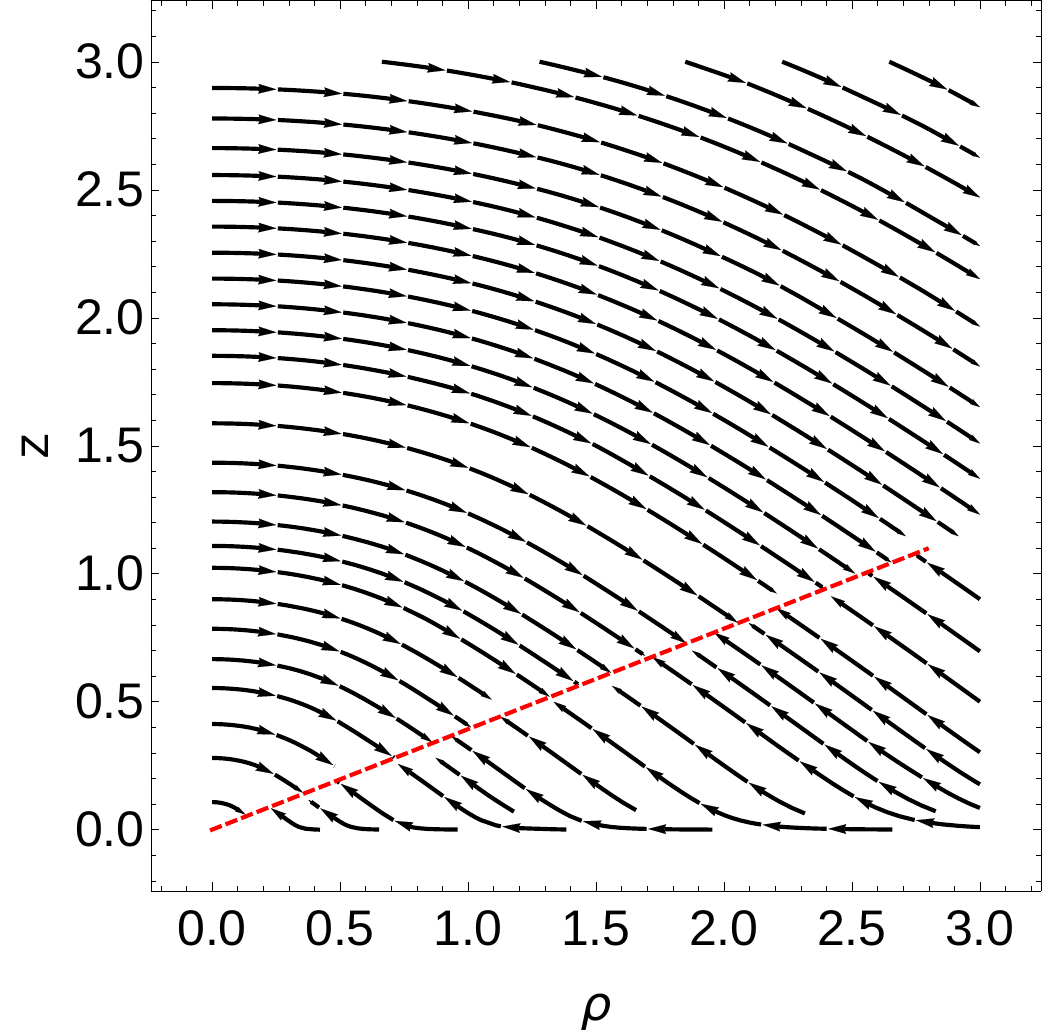}
\put(90,2){(b)}
\end{overpic}
  \begin{overpic}[width=0.32\textwidth]{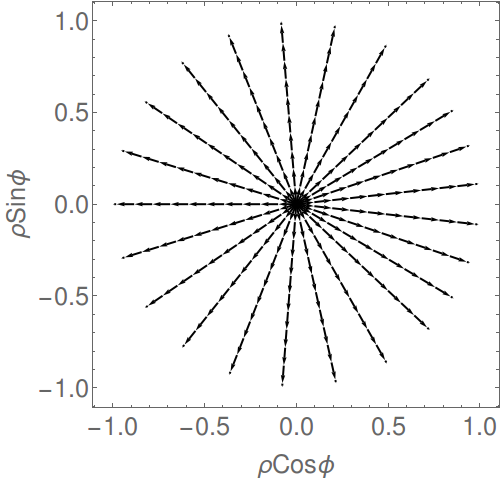}
\put(90,2){(c)}
\end{overpic}
    \begin{overpic}[width=0.32\textwidth]{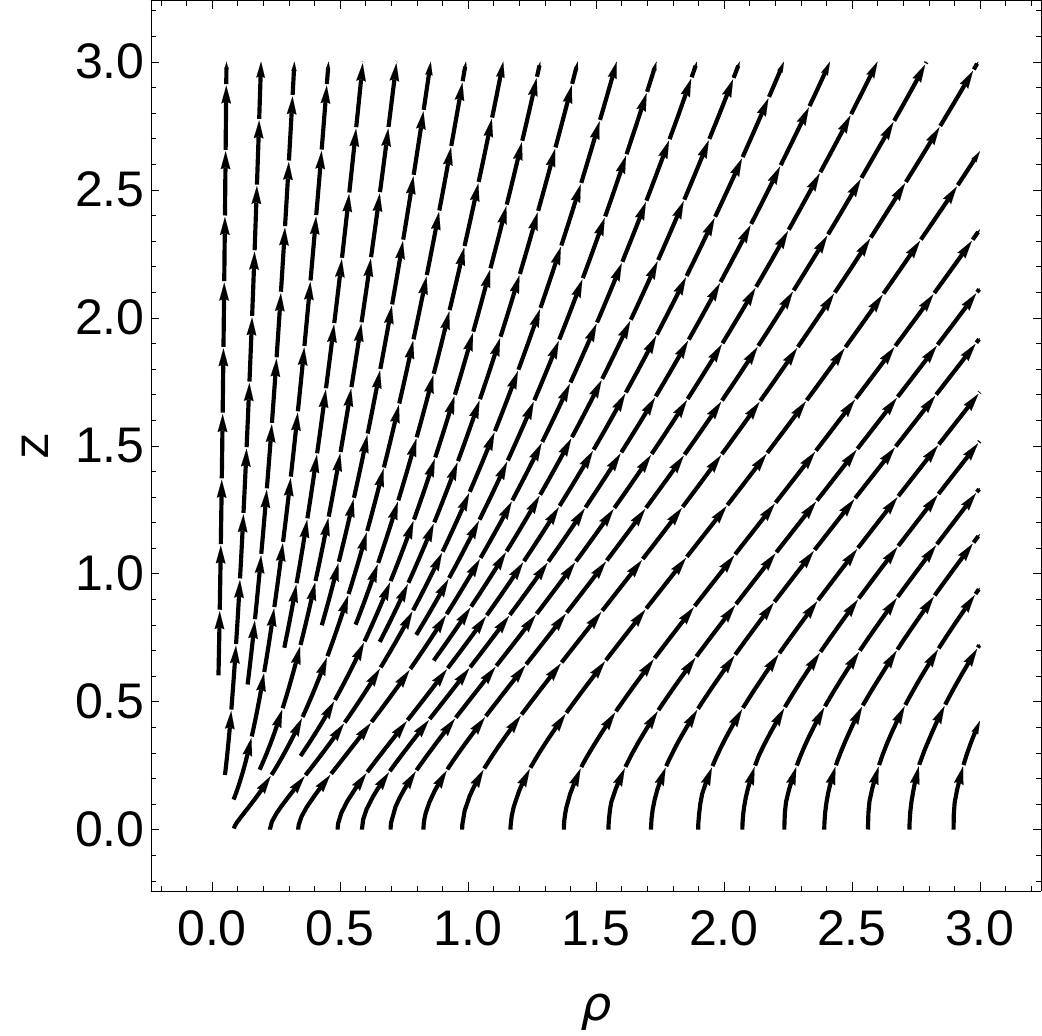}
\put(90,2){(d)}
\end{overpic}
  \begin{overpic}[width=0.32\textwidth]{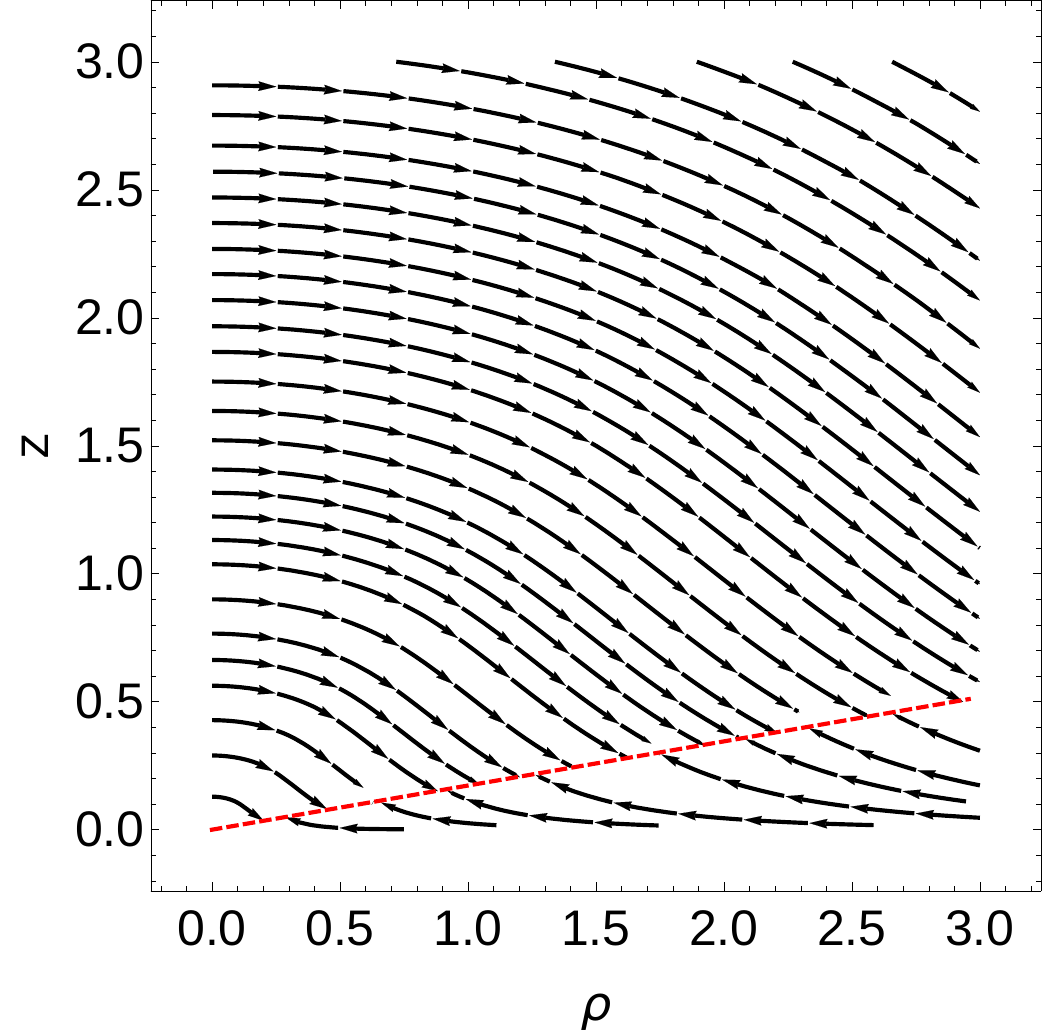}
\put(90,2){(e)}
\end{overpic}
  \begin{overpic}[width=0.32\textwidth]{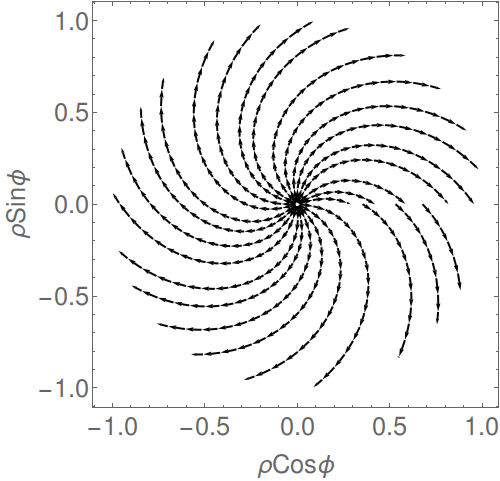}
\put(90,2){(f)}
\end{overpic}
  \caption{Panels (a-c) correspond to the $\xi=0$ solution. (a): Stream lines of the the $B$ field projected onto a vertical plane, with $(\rho,\phi,z)$ being the cylindrical counterparts to the NHEK spatial coordinates. (b): The $E$ field projected onto the same plane. Note that there is an orientation where $|E|$ vanishes, as indicated by the red dashed line. (c): Projection of the $B$ field onto a horizontal plane at $z=0.1$. \\
Panels (d-f) correspond to the $\xi=1$ solution. Respectively, the panels 
(d) (e) and (f) display the analogous information to (a), (b) and (c) for the $\xi=1$ solution. Note in particular that when $\xi \neq 0$, the $B$ field lines acquire a $\phi$ component (f). In panel (f), the streamlines are broken at $\phi=0$, which is a visualization effect due to the streamline integrator working within the $\phi$ range of $[0,2\pi)$. }
	\label{fig:SolLargegStreams}
\end{figure*}

\begin{figure*}[t]
    \begin{overpic}[width=0.99\columnwidth]{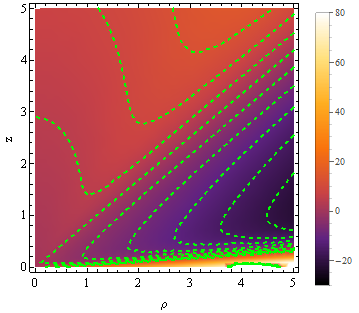}
\put(90,2){(a)}
\end{overpic}
    \begin{overpic}[width=0.99\columnwidth]{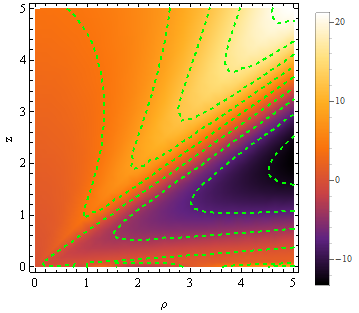}
\put(90,2){(b)}
\end{overpic}
  \caption{(a): For the $\xi=0$ solution. The charge density as $\nabla_a E^a$ for $g_{\text{ex}}$ is plotted on the same vertical half-plane as that used in Fig.~\ref{fig:SolLargegStreams}. Note the contour lines are not equally spaced in values. (b): Similar density plot for the $\xi=1$ solution.}
	\label{fig:ChargeDensity}
\end{figure*}

\begin{figure*}[t,b]
    \begin{overpic}[width=0.85\columnwidth]{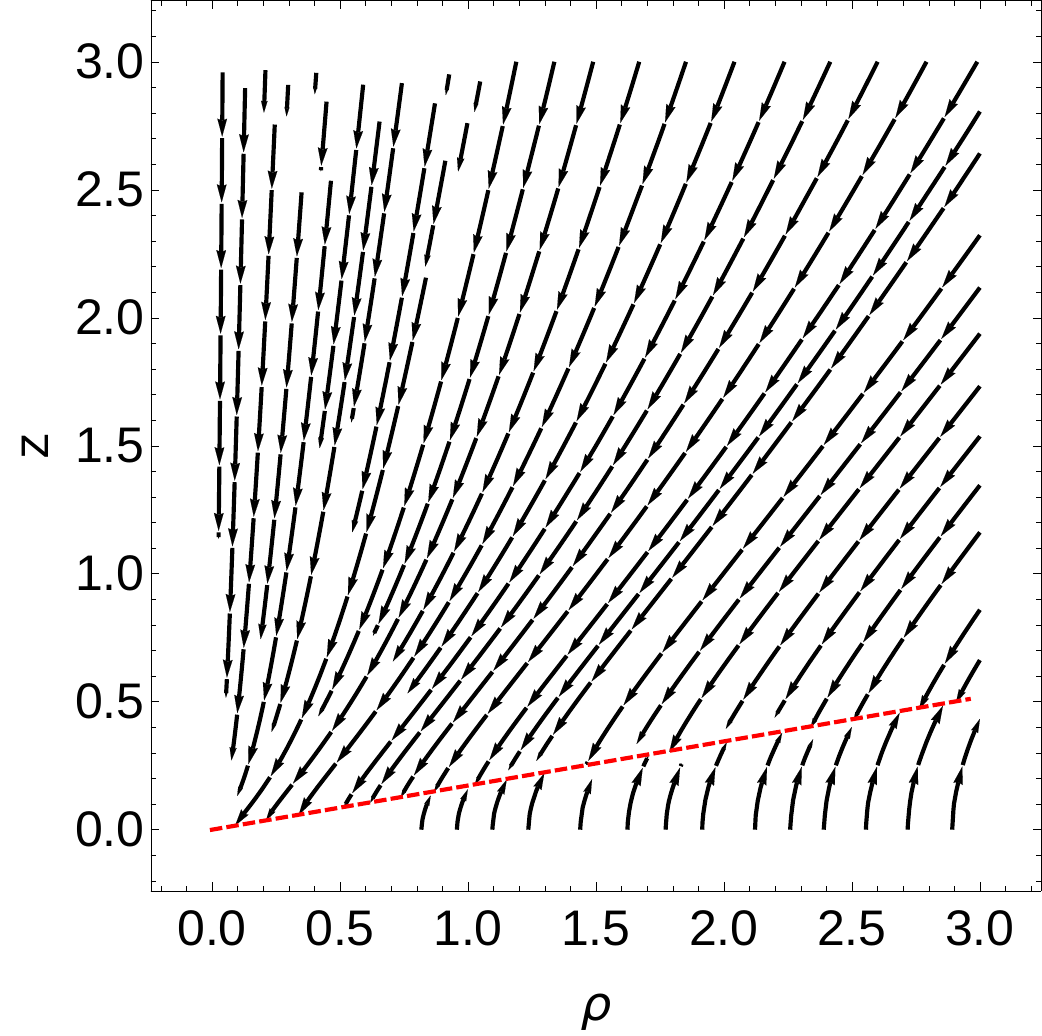}
\put(90,2){(a)}
\end{overpic}
    \begin{overpic}[width=0.99\columnwidth]{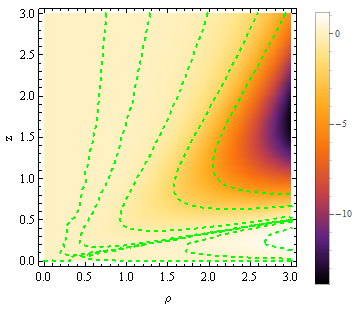}
\put(90,2){(b)}
\end{overpic}
  \caption{(a): For the $\xi=1$ solution. The stream lines of the Poynting vector field projected onto the vertical plane. The red dashed line signifies the location where $|E|$ vanishes. (b): Also for the $\xi=1$ solution. The radial component of the Poynting vector is shown as a density plot on the vertical plane. Note that the contours are not based on equally spaced values. The projections of the Poynting vector for the $\xi=0$ solution vanishes, so there is no corresponding plots for that solution. }
	\label{fig:SolXi1Stream}
\end{figure*}

In this section, we present a couple of non-singular and globally magnetically dominated solutions for $\xi=0$ and $\xi=1$ (with $\alpha=1$). Recall that solutions with larger $g$ magnitudes would be more likely to be magnetically dominated, so we start with $7$ modified Chebyshev basis with an initial coefficient array $\{c_i=1,0,0,0,0,0,0\}$. We then carry out the selective random walk for a moderate amount of trial steps, creating the solutions shown as $g$ in Fig.~\ref{fig:SolLargeg}. We also show reference solutions $g_{\text{ref}}$ (and their residual $\mathcal{L}g_{\text{ref}}$) whose expansion coefficients are proportional to the initial guess but rescaled to have the same $\sum c^2_i$ value as the final outputs of the random walk. We can see that our procedure indeed reduces the residual. However, for the moderate number of random steps allowed, there is still a visible error. 

We then take the $g$ value we obtained at $x=0$ (denoted $g_0$) and impose $g|_{x=0}=g_0$ together with $g'|_{x=0}=0$ as the two boundary conditions, before marching the solution towards $x=1$ using the \verb!NSolve! routine in \verb!Mathematica!. We adjust $g_0$ slightly and arrive at solutions with $g|_{x=1}=-3.7\times 10^{-6}$ for $\xi=0$ and $g|_{x=1}=1.5\times 10^{-7}$ for $\xi=1$, which are shown as $g_{\text{ex}}$ in Fig.~\ref{fig:SolLargeg}. These are our final solutions and we see that $\mathcal{L}g_{\text{ex}}(x)$ are vanishingly small, typical of the output from \verb!NSolve! that did not encounter any problems. In other words, we expect there being no light-surfaces in the interval $[0,1]$. This is indeed the case, which we can see by directly plotting the coefficient of the $g''_{\text{ex}}$ term in $\mathcal{L}g_{\text{ex}}$ (Fig.~\ref{fig:SolLargegLS}), which only nearly, but not exactly, vanishes at $x=0$ (the coefficient evaluates to $0.00553827$ there) for $\xi=0$. We note that the coefficient also vanishes at $x=1$ for both solutions, but this is due to our multiplying a $\Lambda^2$ onto the equation to keep other terms regular, and $x=1$ is not a light surface according to the definition of $|\eta|^2=0$. We verify this by noting that the solutions to $|\eta|^2=0$ are
\bea
g_l^{\pm}=\pm \frac{1}{2}\frac{1+x^2}{\sqrt{1-x^2}}-1, 
\eea
and so the curves $g_l^{\pm}$ should intersect our solutions $g_{\text{ex}}$ at light surfaces. From Fig.~\ref{fig:SolLargegLS}, we see that such an intersection only nearly occurs at $x=0$ for $\xi=0$ and not at $x=1$ for either solution. Our smooth-solution-seeking residual minimization method has thus led us to two solutions without light surfaces. 
We emphasize that the initial guess of $g_0$ provided by the minimization method is of key importance.
Otherwise, the use of arbitrary values for $g_0$ generically leads to light surfaces at which $g$ diverges. 

In Fig.~\ref{fig:SolLargegMagDom}, we plot $B^2-E^2$ for $g_{\text{ex}}$ which shows that both solutions are globally magnetically dominated (as $B^2-E^2$ is a gauge invariant contraction of the Faraday tensor, this conclusion is coordinate/slicing independent), and in Fig.~\ref{fig:SolLargegStreams}, 
we show the projections of $B$ and $E$ fields on a vertical and a horizontal plane, under the cylindrical counterpart to the NHEK coordinates. In addition, the charge density distribution on the vertical plane is shown in Fig.~\ref{fig:ChargeDensity}. 
For readers interested in utilizing these particular solutions, we provide a polynomial fit to $g_{\text{ex}}$, which is 
\bea\label{soln1}
g_{\text{ex}}(x) &\approx&-4.00867 x^7+23.6177 x^6-49.2235 x^5\notag \\
&&+47.8349 x^4-24.007 x^3+6.91724 x^2\notag \\
&&+0.326852 x-1.4573\,,
\eea
for $\xi=0$, and 
\bea\label{soln2}
g_{\text{ex}}(x) &\approx& -1.06308 - 0.181322 x + 5.20729 x^2 \notag \\
&&- 16.5322 x^3 + 41.3352 x^4 - 54.1385 x^5 \notag \\
&& + 32.1822 x^6 - 6.80647 x^7\,,
\eea
for the $\xi=1$ solution. 
We note that when $\xi \neq 0$, the $g/\Lambda$ term in Eq.~\eqref{eq:FaradaySol} can potentially diverge as $x\rightarrow 0$ because $\Lambda \rightarrow 0$, so we plot this quantity in Fig.~\ref{fig:SolLargegMagDom} (b), which shows that $g$ approaches zero faster than $\Lambda$, so all the coefficients in Eq.~\eqref{eq:FaradaySol} remain regular at the poles. 

Lastly, we note that a significant difference between the $\xi=1$ and $\xi=0$ solutions is that the Poynting vector (Eq.~\ref{eq:Poynting}) associated with the former acquires a non-vanishing radial component, signifying energy transfer towards and away from the event horizon. [Notice that for the case $\xi=0$ there is only flux along the $\phi$ direction]. 
In Fig.~\ref{fig:SolXi1Stream} (a), we plot the projection of the Poynting vector onto a vertical plane, and in Fig.~\ref{fig:SolXi1Stream} (b), we plot the radial component of the Poynting vector on the same plane as a density map. As is evident from the figures, the bulk of the energy flux takes
place at a cone centered at around $\theta=64^o$ with an opening of about $20^o$. 

\section{Conclusion}
The FFE equations are a highly non-linear collection of coupled partial differential equations, so in full generality they are very difficult to solve analytically, and are instead usually tackled via numerical simulations. Nevertheless, one can concentrate on situations possessing a high degree of symmetry, which allows for simplifying the equations and obtaining semi-analytical (only certain simpler components of the overall solution are obtained numerically) solutions. For example, when the background metric is stationary and axisymmetric, one can require that the FFE solution to also respect these symmetries, in which case the Euler potentials can be written in terms of a few functions with highly restricted forms. These functions are the magnetic flux function $\psi$, the polar current $I$ and the angular velocity of the field lines $\Omega_F$. The prescription of convenient choices of $I$ and $\Omega_F$ that leads to simplifications of the final stream equation (that determines $\psi$) is a major step towards obtaining semi-analytical solutions. 

In this paper, we fix these quantities by imposing a restricted form of a third symmetry. Namely we work inside the NHEK spacetime that possesses a conformal symmetry, and demand that the field tensor -- as well as the Euler potentials -- of our FFE solution be self-similar under the associated transformations. This fixes the functional forms of $I$ and $\Omega_F$ in terms of $\psi$, and reduces the stream equation to a single second order ODE. Due to the existence of light surfaces on which the steam equation becomes locally first order, it is difficult to find non-singular solutions using traditional ODE-solving techniques. Accordingly, we have developed a residual minimization method tailored to the task of finding regular solutions. We have also shown that using this method, we can find non-singular solutions that are globally magnetically dominated (thus physically realistic). 
Our study therefore complements earlier works that have found partially electrically dominated solutions in NHEK, and lays the necessary groundwork for systematically generating further FFE solutions.
 The complexity-reducing procedure as well as the
technique for solving the resulting equation through residual minimization, as employed here, should also be applicable 
to other FFE problems.

The aim of this paper has been to introduce the fundamental equations and methods. In order not to overly-complicate the discussion, we have only described two specific solutions as demonstrative examples. (Notice however that 
further ones related by $\text{SL}(2,{\Bbb R})$ transformations of these can be straightforwardly obtained as 
discussed in~\cite{Lupsasca:2014pfa}).
We will leave the production and examination of additional interesting semi-analytical solutions to future works. In particular, the solutions we obtained map to time-dependent near-horizon solutions of near-extremal Kerr black holes. It remains interesting to explore possible time-stationary solutions, especially the ones with power extraction from the horizon (see the discussions about black hole ``Meissner effect" in \cite{PhysRevD.10.1680,King1975,Penna:2014aza}). 
We expect future studies building on the foundations presented in this paper to further illuminate the properties of magnetospheres in the near-horizon region of rapidly rotating black holes. 

\acknowledgements
We thank Alexandru Lupsasca and Ted Jacobson for insightful discussions. This work was supported by NSERC through a Discovery Grant (to LL) and CIFAR (to LL). FZ would like to thank Perimeter Institute for hospitality during the very early stages of this work. 
Research at Perimeter
Institute is supported through Industry Canada and by the Province of Ontario
through the Ministry of Research \& Innovation.  

\bibliography{FFE_NHEK_prd.bbl} 

\end{document}